# Uncertainty-Aware Genomic Classification of Alzheimer's Disease: A Transformer-Based Ensemble Approach with Monte Carlo Dropout

Taeho Jo[a]*, Eun Hye Lee[a] and Alzheimer's Disease Sequencing Project

[a] Indiana Alzheimer Disease Research Center and Center for Neuroimaging, Department of Radiology and Imaging Sciences, Indiana University School of Medicine, Indianapolis, IN 46202, USA

**Corresponding authors**

Taeho Jo, PhD*

Indiana Alzheimer Disease Research Center and Center for Neuroimaging, Department of Radiology and Imaging Sciences, Indiana University School of Medicine, Indianapolis, IN 46202, USA

Tel.: +1-317-963-7504

E-mail: tjo@iu.edu

**Abstract**

**INTRODUCTION:** Alzheimer's disease (AD) is genetically complex, complicating robust classification from genomic data.

**METHODS:** We developed a transformer-based ensemble model (TrUE-Net) using Monte Carlo Dropout for uncertainty estimation in AD classification from whole-genome sequencing (WGS). We combined a transformer that preserves single-nucleotide polymorphism (SNP) sequence structure with a concurrent random forest using flattened genotypes. An uncertainty threshold separated samples into an uncertain (high-variance) group and a more certain (low-variance) group.

**RESULTS:** We analyzed 1050 individuals, holding out half for testing. Overall accuracy and area under the receiver operating characteristic (ROC) curve (AUC) were 0.6514 and 0.6636, respectively. Excluding the uncertain group improved accuracy from 0.6263 to 0.7287 (10.24% increase) and F1 from 0.5843 to 0.8205 (23.62% increase).

**DISCUSSION:** Monte Carlo Dropout–driven uncertainty helps identify ambiguous cases that may require further clinical evaluation, thus improving reliability in AD genomic classification.

**Abbreviations**

- AD: Alzheimer's disease
- TrUE-Net: Transformer-based, Uncertainty-aware, Ensemble Network
- SNP: single-nucleotide polymorphism
- WGS: whole-genome sequencing
- ROC: receiver operating characteristic
- AUC: area under the ROC curve
- F1: F1-score

**Introduction**

Alzheimer's disease (AD) is a complex, progressive neurodegenerative disorder and the most common cause of dementia in the elderly. It represents a significant public health concern as populations age, with tens of millions affected worldwide. From a genetic standpoint, AD has a highly polygenic architecture: while rare familial early-onset AD may result from single-gene mutations (e.g. in *APP*, *PSEN1*, *PSEN2*) the more common late-onset AD involves dozens of risk-associated genetic variants, each conferring a modest effect (Bekris et al., 2010; Saykin et al., 2015). Twin studies estimate AD heritability at approximately 70%,(Gatz et al., 1997; M. Gatz et al., 2006) yet genome-wide association studies (GWAS) have collectively identified over 80 loci (including more than 100 independent variants) that influence AD risk (Andrews et al., 2023; Bellenguez et al., 2022; Jansen et al., 2019; Kunkle et al., 2019). This genetic complexity, along with gene–gene and gene–environment interactions, makes genomic variant classification for AD prediction challenging. Individual variants typically have limited predictive power, and there is substantial "missing heritability," indicating that many genetic factors and their effects remain uncertain.

Traditional statistical models and polygenic risk scores have had limited success in AD prediction due to the high-dimensional, often noisy nature of genomic data.(Genomes Project et al., 2015) Recently, researchers have turned to deep learning to capture nonlinear interactions among variants and improve predictive accuracy. For example, fully-connected neural networks have been applied to AD genomics data, reporting moderate accuracy (~68-70%) in distinguishing AD dementia from controls (Jo et al., 2023; Shigemizu et al., 2023). Transformers, a deep learning architecture originally developed for natural language processing, have shown notable ability to model long-range dependencies in sequential data and have been adapted to genomic sequences (Jo et al., 2025).

While deep learning can yield improved accuracy, the reliability of predictions remains a major concern in medical applications. Black-box models often provide outputs without clearly indicating confidence levels, and can exhibit overconfidence even when they are incorrect (Milanés-Hermosilla et al., 2021). This issue is particularly critical in clinical genomics, where misclassification of a patient's risk may lead to inappropriate interventions and reduce user trust (Ferrante et al., 2024; Nguyen et al., 2021). To address these challenges, we propose an uncertainty-aware deep learning framework for AD genomic variant classification that leverages Monte Carlo Dropout. We refer to this approach as TrUE-Net (Transformer-based, Uncertainty-aware, Ensemble Network). Monte Carlo Dropout is a Bayesian approximation technique that keeps dropout active at inference time to generate multiple stochastic forward passes, enabling an estimation of predictive variance.

A notable aspect of the proposed method is its integration of a transformer-based genotype classifier with a concurrently trained random forest model. Their respective probability outputs are aggregated through a learned weighting factor, thereby producing a single combined probability estimate. By merging the transformer's dropout-induced variance with the random forest's estimator variance, the framework yields a comprehensive measure of predictive uncertainty for each sample. This unified uncertainty metric facilitates the identification of low-confidence predictions that warrant additional scrutiny, thereby enhancing overall reliability. Empirical evaluations indicate that excluding high-variance predictions can significantly improve classification performance on the remaining, more confidently assessed subset, providing a practical strategy for mitigating diagnostic risk in Alzheimer's disease genomics.

**Methods**

**Dataset and Genomic Preprocessing**

We evaluated our approach using whole-genome sequencing (WGS) data from a total of 1,566 participants, drawn from the Alzheimer's Disease Neuroimaging Initiative (ADNI). Of these, 516 individuals diagnosed with mild cognitive impairment were excluded, resulting in a final case-control set of 443 cognitively normal (CN) participants and 607 with clinically confirmed AD dementia (Table1). Diagnostic criteria for ADNI participants are provided on the ADNI website (http://www.adni-info.org). Sequencing was performed on Illumina platforms with either 100-bp or 150-bp paired-end reads aligned to the GRCh38 (hg38) reference genome using BWA-MEM,(Li & Durbin, 2009) followed by PCR duplicate marking, local realignment, and base quality score recalibration with the Genome Analysis Toolkit (GATK). Joint variant calling proceeded via GATK HaplotypeCaller, and the Genome Center for Alzheimer's Disease (GCAD) applied its Variant Calling Pipeline for standardized quality control, encompassing checks of single-nucleotide variant (SNV) concordance, sex mismatches, relatedness, and contamination.

Sample-level quality control removed individuals exhibiting sex inconsistencies, call rates below 95%, or duplicated genetic profiles, and excluded those with high relatedness (Pihat > 0.4). Variant-level filtering discarded SNPs with call rates under 95%, Hardy–Weinberg equilibrium p-values less than $1 \times 10^{-6}$, or minor allele frequencies below 1%. Variants that appeared monomorphic, multi-allelic, or had poor genotype quality (GQ < 20), low read depth (DP < 10), or missingness exceeding 20% were also set aside. In parallel, population structure was implicitly managed because both ADNI and ADSP-FUS1-ADNI-WGS-2 primarily included participants of European ancestry, and outliers had already been removed during each study's enrollment. We then applied a k-nearest neighbors (k-NN) imputation method (Jo et al., 2025) to address residual missing genotypes following these QC steps.

After these procedures, approximately 100,000 common variants per individual—including those encompassing the APOE locus—were retained for the final analyses. Each variant was encoded as an integer genotype (0, 1, or 2). We split the dataset into equally sized training and test sets, ensuring stratification by diagnostic label. We partitioned the dataset into training and test subsets, using a 5-fold stratified cross-validation on the training set to monitor variability and validate the chosen hyperparameters. Final assessments of generalization were subsequently carried out on the hold-out test set.

**An Ensemble Approach with Window-Based and Flattened Genomic Representations**

We constructed a transformer-based classification model within the proposed TrUE-Net framework to distinguish AD dementia cases from CN controls by translating each individual's genomic data into a structured sequence. In this approach, we segmented each individual's array of single-nucleotide polymorphisms (SNPs) into non-overlapping tokens of length 100 to balance resolution and computational feasibility. These token-level representations were subsequently projected into a feature dimension of 128 through a linear embedding layer, followed by ReLU activation and dropout at a rate of 0.2.

Subsequently, a stack of two transformer encoder layers processed the embedded windows in order to capture relevant dependencies among the segments. Each encoder layer implemented four attention heads and employed a feed-forward sublayer of dimension 512, allowing the model to attend to both local interactions and more distant correlations across the genotypic windows. After the transformer operations, the representations of all windows for a given individual were averaged to form a single feature vector. This pooled vector was then fed into a linear layer with two output units, whose logits were trained to discriminate AD dementia from CN controls. We minimized the cross-entropy loss function to fit this model, and we applied a softmax function during inference to interpret the logits as class probabilities. Throughout this pipeline, Monte Carlo Dropout remained active, enabling multiple stochastic forward passes for each sample at test time and facilitating the estimation of predictive variance.

In parallel, we trained a random forest classifier to provide an alternative perspective on the same genomic data. Instead of preserving any sequential arrangement of the SNP windows, this second model operated on a concatenated vector comprising the entire genotype profile, effectively flattening all windows into a single feature array. This arrangement treated each variant position as an independent predictor, disregarding explicit local ordering but potentially capturing broad, global patterns through the aggregate of decision trees. To coordinate these two complementary classification strategies, we ensured that both models were trained and evaluated on the same set of participants, which allowed subsequent ensemble weighting.

Within the training set, we carried out a five-fold cross-validation procedure to identify optimal hyperparameters, including the weighting factor that merged outputs from the transformer and random forest models, along with a variance threshold to distinguish uncertain predictions. Specifically, we averaged the best weighting factors and threshold levels discovered across the five folds to arrive at the final parameter settings. We subsequently retrained both the transformer-based model and the random forest on the entire training set, using an AdamW optimizer with a learning rate of 0.001 and a batch size of 32 over five epochs. Throughout the training, the dropout-driven variance was monitored to encourage stable latent representations while preserving the stochasticity needed for uncertainty estimation.

Lastly, we applied the combined models to the test set to gauge the final predictive performance. For each individual in the test set, we computed a weighted average of the transformer's probabilistic output and the random forest's prediction, using the previously determined combination weight. Similarly, we merged the corresponding variance estimates to assign each sample either to an uncertain or a more confidently assessed category, depending on whether its variance exceeded or fell below the threshold observed in cross-validation. This procedure enabled us to calculate separate metrics for the uncertain and more certain groups and provided insight into the reliability of the resulting ensemble model. By segregating test samples in this manner, we aimed to highlight the potential benefits of uncertainty quantification in identifying cases that may require additional clinical follow-up or expert review.

**Predictive Variance for Uncertainty Estimation in AD Classification**

Within the training set, a five-fold cross-validation procedure was carried out to generate out-of-fold (OOF) predictions from both a transformer-based model with Monte Carlo Dropout and a RandomForest classifier, facilitating the discovery of key hyperparameters: the ensemble weight α and a variance threshold to distinguish uncertain from more confidently classified samples. During training, the following variance penalty was added to the cross-entropy objective:

$$\mathcal{L} = \text{CrossEntropy}(\hat{y}, y) + \lambda_{unc} \cdot \text{mean}(\text{Var}(\text{logits}))$$

Where Var(logits) quantifies the dropout-induced variance in the logit outputs over multiple forward passes, and $\lambda_{unc}$ (for instance, 0.05) encourages the model to avoid excessive fluctuations. At inference, Monte Carlo Dropout remained active, producing an average probability $P_{trans}$ and a per-sample variance $Var_{trans}$.

A RandomForest classifier was trained in parallel on a flattened representation of the same SNP data, providing an independent probability, $P_{rf}$, and an uncertainty measure, $Var_{rf}$, based on the variance in its aggregate of decision trees. To formulate a single matrix expression that merges these two models, one can define the following 2×2 matrix and 1×2 vector:

$$M = \begin{pmatrix} P_{trans} & Var_{trans} \\ P_{rf} & Var_{rf} \end{pmatrix}, \quad w = (\alpha \quad 1 - \alpha)$$

where α was selected to maximize accuracy on the OOF predictions. The final ensemble output,

$$O_{ensemble} = w \cdot M$$

thus produces a 1×2 vector whose first component is the ensemble probability $P_{ensemble}$ and whose second component is the ensemble variance $Var_{ensemble}$. An extensive grid of candidate thresholds on $Var_{ensemble}$ was evaluated (for instance, by partitioning the observed range of variance values into multiple segments), and the threshold that optimized a chosen metric (e.g., the sum of AUCs in uncertain and more certain subsets) was recorded in each fold. Once all folds were completed, their best hyperparameters were averaged to finalize α and the variance threshold.

After collecting the OOF predictions, both models were retrained on the entire 50% training setusing the final values of α and the variance threshold. Monte Carlo Dropout remained active for the transformer in order to continue providing a mean probability and variance for each sample. Finally, the refitted models were applied to the unseen 50% test set for an unbiased evaluation of predictive performance. Each test sample's ensemble variance was compared to the threshold, and any sample above that threshold was classified as uncertain group, while the rest were deemed relatively certain. This procedure facilitated an assessment of whether variance-based filtering effectively identified samples that were more prone to misclassification, by comparing accuracy and related metrics across uncertain versus more confidently classified subsets.

The source code for this study is available at GitHub: https://github.com/taehojo/TrUE-Net

**Results**

A total of 525 test samples were used after training the TrUE-Net model on the remaining data. The final predictive run across these 525 samples, referred to here as the All group, achieved an overall accuracy of 0.6514, an area under the ROC curve of 0.6636, and an F1-score of 0.6679. Figure 1 presents the test samples that divided into an uncertain group and a certain group by applying a threshold to this score.Where the horizontal axis represents the predicted probability of AD dementia and the vertical axis shows the model's variance estimate. 396 of the test samples, corresponding to 75.4 percent, were labeled uncertain, while 129, corresponding to 24.6 percent, were categorized as certain. Figure 2 shows accuracy, area under the ROC curve, and F1 metrics for both groups. The uncertain group had values of 0.6263, 0.6268, and 0.5843 for these measures, while the certain group reached 0.7287, 0.6816, and 0.8205. Figure 3 shows that uncertain samples cluster near the 0.3 to 0.5 range, whereas certain samples appear in more definitive ranges below 0.3 or above 0.6.

To optimize the relative weighting of the transformer and random forest outputs, multiple values of $\alpha$ were evaluated in increments of 0.2, from 0 to 1. The highest AUC (0.6508) emerged at $\alpha = 0.6$, whereas $\alpha = 1$ yielded the lowest AUC (0.6283). Intermediate values (e.g., $\alpha = 0.4$) resulted in slightly lower AUCs (0.6467) compared to the peak at $\alpha = 0.6$. In parallel, a threshold parameter was varied to categorize each sample into uncertain versus certain subgroups. Ten candidate thresholds were selected by dividing the range between the minimum and maximum uncertainty values into equal intervals. Each threshold was then used to compute a performance score reflecting the sum of the AUC values for the uncertain and certain subsets combined. As shown below, the highest combined score was achieved at a threshold of 0.0741.

Figure 4 compares two histograms for each diagnosis category: CN and AD dementia. Figure 4(A) shows the final predicted probability of AD dementia. Here, the CN group exhibits a pronounced peak near 0.4, whereas the AD dementia group's distribution shifts to approximately 0.6. Despite some overlap around 0.4 to 0.6, CN samples are generally more prevalent in the lower-probability range, and AD dementia samples concentrate in the higher-probability range. In Figure 4(B), the same two groups are compared in terms of model uncertainty. The CN group reaches its highest peak near 0.10, whereas the AD dementia group's distribution is somewhat centered around 0.07 to 0.09. Although these peaks suggest a slight difference between CN and AD dementia, both groups occupy similar variance ranges overall.

## Discussion

This study investigated an uncertainty-aware genomic classification framework for AD by employing a transformer-based model with Monte Carlo Dropout. Our TrUE-Net approach set variance-based thresholds to single out test samples associated with higher predictive variance, termed uncertain, from those deemed more stable, termed certain. This separation provided a way to focus on reliably classified samples while drawing attention to ambiguous cases in need of additional diagnostic steps. Moreover, performance improved considerably for the certain group. These findings support earlier work suggesting that a relatively small percentage of ambiguous genotypes can contribute disproportionately to overall errors, and that removing them from the main classification pipeline can yield marked gains in core metrics(Belloy et al., 2022; Escott-Price et al., 2019; Jo et al., 2022).

The final prediction results of All group presented moderate performance. This indicated that an all-or-nothing scenario does not fully exploit the advantages of variance-based filtering. However, an additional analysis involving predictive variance revealed substantial differences in performance between subsets of samples with high variance and those with low variance. Although improved performance on a smaller subset is a recognized effect of uncertainty-driven approaches, the degree of improvement here is notable. The uncertain group's lower accuracy illustrated the complexity introduced by higher predictive variance. Probability estimates for these samples often hover near intermediate values, making the classification boundary less clear. Notably, these uncertain samples are distributed across both CN and AD, indicating that the model's predictive variance is not restricted to a single group. Identifying these ambiguous cases allows clinicians or researchers to allocate additional diagnostic steps, potentially reducing diagnostic errors.

In a separate analysis that complements these results, excluding predictions tagged with high predictive variance elevated accuracy and increased the area under the ROC curve. This observation demonstrates how ignoring or postponing ambiguous classifications can lead to tangible benefits. The method also allows for more nuanced output than a simple binary label by providing an estimate of confidence through variance. This capacity can be valuable in clinical contexts, where errors carry weighty consequences, since higher-variance cases can receive extra diagnostic attention such as neuroimaging or biomarker testing. Conversely, lower-variance classifications may be handled with more confidence, potentially increasing the efficiency of clinical workflows.

Nevertheless, several limitations remain. First, the dataset was primarily composed of 1050 individuals of European descent, which constrains generalization to other ancestral groups. Second, because mild cognitive impairment cases were omitted, the analysis offered only a binary comparison between individuals without cognitive decline and those with established Alzheimer's dementia. This design does not capture subtle genetic features linked to early or transitional stages, and it does not incorporate gene interactions often observed in complex disorders. Third, the model architecture did not address rare variants that can strongly affect disease risk. Lastly, although variance-based filtering improved performance on the less ambiguous subset, it reduced the total number of samples that could be confidently classified, raising the need to balance completeness against the aim of higher accuracy and F1.

Further efforts may tackle these challenges by recruiting more diverse populations and including mild or preclinical states, enabling a broader evaluation of how variance-based methods perform across different stages of Alzheimer's disease. It would also be instructive to incorporate additional data types, such as transcriptomics, proteomics, or metabolomics, in order to capture a richer biological background. Methods that dynamically tune the variance threshold could reduce the proportion of excluded samples while sustaining performance gains. Taken together, these avenues can refine transformer-based models for genomic risk prediction and strengthen confidence in their utility in clinical scenarios.

## Conclusion

In this study, we applied TrUE-Net, a transformer-based classification model with Monte Carlo Dropout to estimate uncertainty in the genomic prediction of Alzheimer's disease. Filtering out samples with elevated variance in their predicted probabilities led to higher accuracy and area under the ROC curve on the retained subset, thereby demonstrating the practical benefits of an uncertainty-aware approach. These results underscore the usefulness of explicitly modeling uncertainty for genomic classification tasks, where identifying ambiguous cases can direct further testing or evaluations. The proposed pipeline can be integrated into existing diagnostic workflows for AD and provides a foundation for future research aimed at incorporating other data modalities, larger and more diverse populations, and more advanced uncertainty estimation methods.

## Conflict of Interest Statement

Dr. Taeho Jo serves as an Associate Editor of *Genomics & Informatics*, a Springer-Nature Journal, but had no role in the editorial assessment or peer review of this manuscript. The authors declare no other conflicts of interest.

**Funding**

This research was partially supported by the Alzheimer's Association (AA) under the grant number AARG 22-974053 and the National Institutes of Health (NIH): P30 AG072976, U19 AG024904, U01 AG068057, U19 AG074879, U01 AG072177, U24 AG074855.

**Consent Statement**

Consent from human subjects was not necessary for this study.

| Diagnosis Group | Sample Size | Mean Age (SD) | % Males / Females | APOE ε2/ε2 | APOE ε2/ε3 | APOE ε2/ε4 | APOE ε3/ε3 | APOE ε3/ε4 | APOE ε4/ε4 | APOE ε4 Carriers (%) |
|---|---|---|---|---|---|---|---|---|---|---|
| Cognitive normal (CN) | 443 | 72.63 (6.31) | 53.05 / 46.95 | 2 | 57 | 7 | 257 | 110 | 10 | 27.09 |
| Alzheimer's disease (AD) | 607 | 73.97 (7.25) | 42.01 / 57.99 | 1 | 19 | 14 | 194 | 278 | 101 | 62.44 |

**Table 1.** This table shows demographic and APOE genotype data for the cognitively normal and Alzheimer's disease groups. Among the 443 cognitively normal individuals, the mean age was 72.63 years, 53.05 percent were male, and 27.09 percent carried the APOE ε4 allele. Among the 607 individuals with Alzheimer's disease, the mean age was 73.97 years, 42.01 percent were male, and 62.44 percent were ε4 carriers, consistent with the well-known association of APOE ε4 with higher Alzheimer's disease risk.

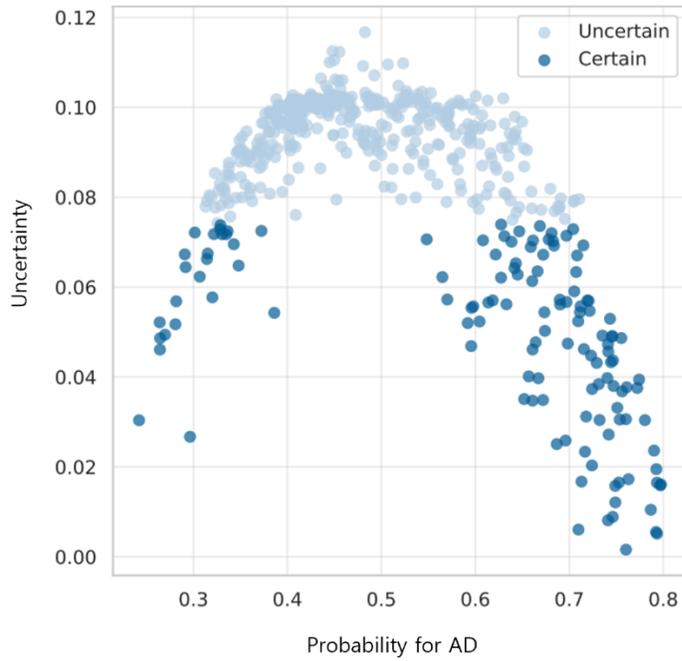

**Figure 1.** This scatter plot compares the predicted probability for Alzheimer's disease on the horizontal axis with the model's uncertainty on the vertical axis. Lighter points labeled uncertain generally exhibit higher uncertainty, while darker points labeled certain show lower uncertainty.

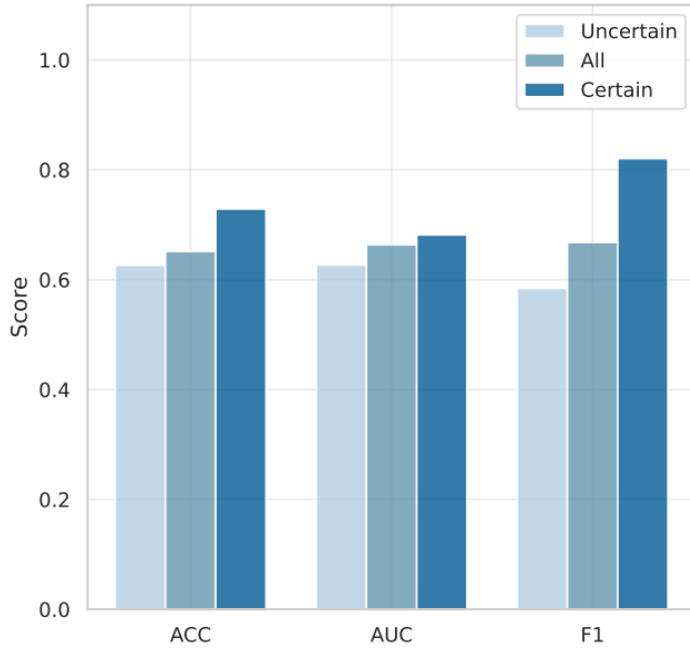

**Figure 2.** This bar chart displays accuracy, area under the ROC curve, and F1-score for three subsets: uncertain, shown with the lightest bars, all test samples, shown with medium bars, and certain, shown with the darkest bars. The uncertain subset has lower scores overall, while the certain subset achieves higher values on each metric.

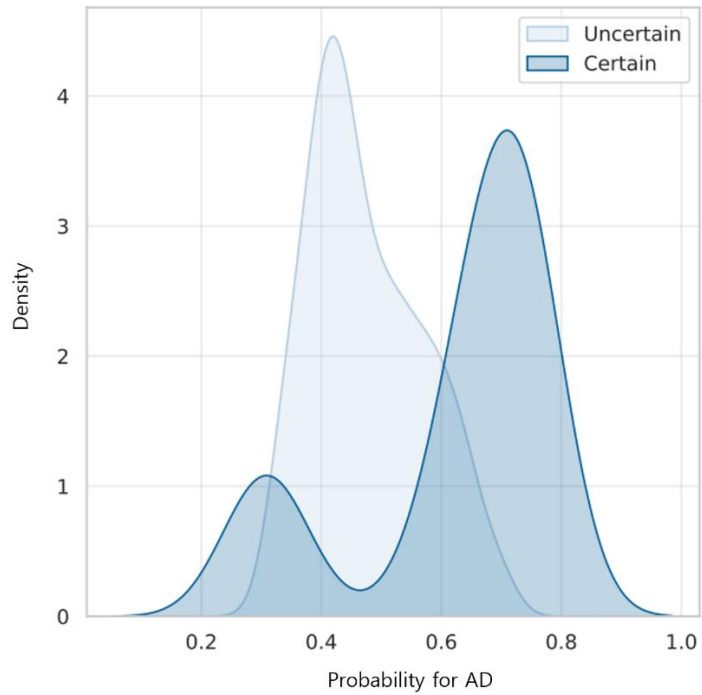

**Figure 3.** This kernel density plot displays the distribution of predicted probabilities for uncertain in a lighter color and certain in a darker color. The uncertain group clusters near mid-range values around 0.3 to 0.5, whereas the certain group appears at more extreme probabilities below 0.3 or above 0.6.

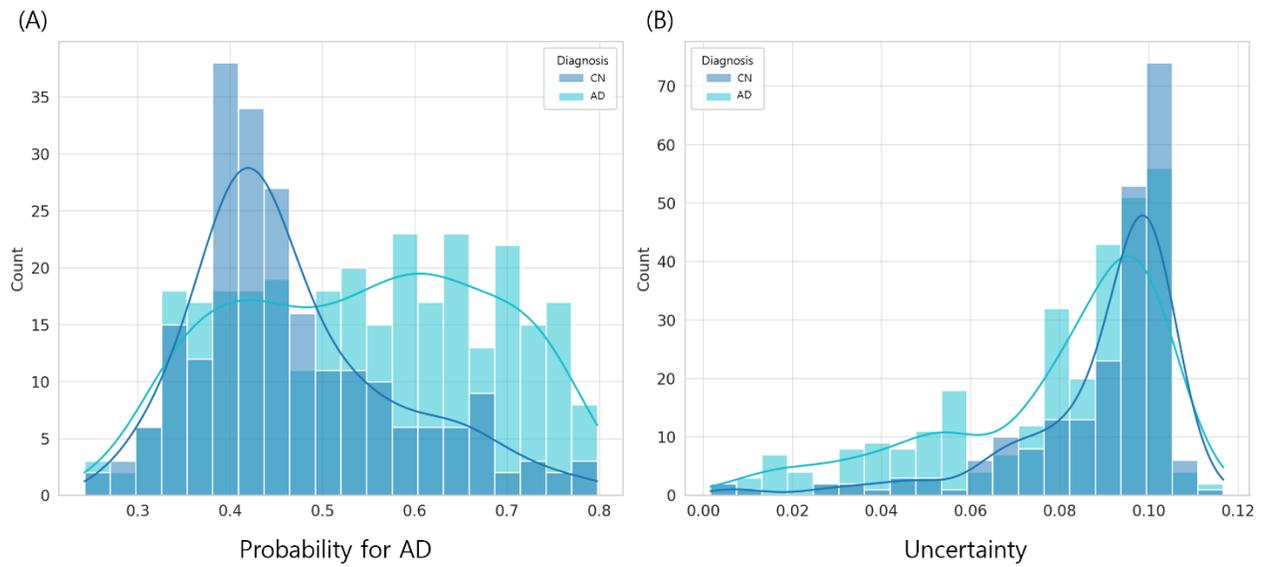

**Figure 4.** (A) compares the distribution of final predicted probabilities for the cognitively normal (CN) and AD groups, showing that CN cases tend to peak near 0.4 while AD cases shift toward higher probabilities around 0.6. (B) shows the distribution of model uncertainty, where CN reaches its highest frequency near 0.10 and AD clusters slightly below that range, although substantial overlap remains between the two groups.